\documentclass[12pt]{article}
\usepackage{amssymb,latexsym}
\def\ZZ{{\mathbb Z}}
\def\QQ{{\mathbb Q}}
\def\RR{{\mathbb R}}

\newtheorem{formula}{}[section]
\newtheorem{proposition}[formula]{Proposition}
\newtheorem{definition}[formula]{\indent Definition}
\newtheorem{corollary}[formula]{\indent Corollary}
\newtheorem{remark}[formula]{\indent Remark}
\newtheorem{lemma}[formula]{\indent Lemma}
\newtheorem{theorem}[formula]{\indent Theorem}

\def\thrm{\begin{theorem}}
\def\thrml#1{\begin{theorem}\label{#1}}
\def\ethrm{\end{theorem}}
\def\rmrk{\begin{remark}}
\def\rmrkl#1{\begin{remark}\label{#1}}
\def\ermrk{\end{remark}}
\def\dfntn{\begin{definition}}
\def\dfntnl#1{\begin{definition}\label{#1}}
\def\edfntn{\end{definition}}
\def\nmrt{\begin{enumerate}}
\def\enmrt{\end{enumerate}}

\def\qtn{\begin{equation}}
\def\qtnl#1{\begin{equation}\label{#1}}
\def\eqtn{\end{equation}}
\def\lmm{\begin{lemma}}
\def\lmml#1{\begin{lemma}\label{#1}}
\def\elmm{\end{lemma}}
\def\crllr{\begin{corollary}}
\def\crllrl#1{\begin{corollary}\label{#1}}
\def\ecrllr{\end{corollary}}

\begin{document}
\title{}
\date{}
\maketitle
\vspace{-0,1cm}
\centerline{\bf TROPICAL DIFFERENTIAL EQUATIONS}
\vspace{7mm}
\author{
\centerline{Dima Grigoriev}
\vspace{3mm}
\centerline{CNRS, Math\'ematique, Universit\'e de Lille, Villeneuve
d'Ascq, 59655, France} \vspace{1mm} \centerline{e-mail:\
dmitry.grigoryev@math.univ-lille1.fr } \vspace{1mm}
\centerline{URL:\ http://en.wikipedia.org/wiki/Dima\_Grigoriev} }

\begin{abstract}
Tropical  differential equations are introduced and an algorithm is
designed which tests solvability of  a system of tropical linear
differential equations within the complexity polynomial in the size
of the system and in its coefficients. Moreover, we show that there
exists a minimal solution, and the algorithm constructs it (in case
of solvability). This extends a similar complexity bound established
for tropical linear systems. In case of tropical linear differential
systems in one variable a polynomial complexity algorithm for
testing its solvability is designed.

We prove also that the problem of solvability of a system of
tropical non-linear differential equations in one variable is
$NP$-hard, and this problem for arbitrary number of variables
belongs to $NP$. Similar to tropical algebraic equations, a tropical
differential equation expresses the (necessary) condition on the
dominant term in the issue of solvability of a differential equation
in power series.
\end{abstract}

{\bf Keywords}: tropical differential equations, polynomial complexity
solving

\section*{Introduction}

Tropical algebra deals with the tropical semi-rings $\ZZ_+$ of
non-negative integers or $\ZZ_+\cup \{\infty\}$ endowed with the
operations $\{\min, +\}$, or with the tropical semi-fields $\ZZ$ or
$\ZZ \cup \{\infty\}$ endowed with the operations $\{\min, +, -\}$
(see e.~g. \cite{I}, \cite{M}, \cite{T}).

A {\it tropical linear differential equation} is a tropical linear
polynomial of the form
\begin{eqnarray}\label{1}
\min_{i,j} \{a_i^{(j)} + x_i^{(j)},\, a\}
\end{eqnarray}
where the coefficients $a,\, a_i^{(j)} \in \ZZ_+ \cup \{\infty\}$,
and a variable $x_i^{(j)}$ is treated as "$j$-th derivative of
$x_i:=x_i^{(0)}$".

For a subset $S_i\subset \ZZ_+$ we define the {\it valuation}
$$Val_{S_i}(\{j\ge 0\}):= Val_{S_i}(\{x_i^{(j)}\}_{j\ge 0}):\ZZ_+ \to
\ZZ_+\cup \{\infty\}$$ \noindent of variable $x_i$ as follows. For
each $j\ge 0$ take the minimal $s\in S_i$ (provided that it does
exist) such that $s\ge j$ and put $Val_{S_i}(j):=s-j$: in case when
such $s$ does not exist put $Val_{S_i}(j):=\infty$. We use a
shorthand $$Val_{S_1,\dots,S_n}:=Val_{S_1}\times \cdots \times Val_{S_n}: \ZZ_+^n \to
(\ZZ_+ \cup \{\infty\})^n.$$

Observe that if $X_i$ is a power series in $t$ with the support
$\{t^s,\, s\in S_i\}$ then $Val_{S_i} (j)$ is the order $ord_t
(X_i^{(j)})$ at zero of the $j$-th derivative $X_i^{(j)}$.

We say that $S_1,\dots, S_n$ is a {\it solution of the  tropical linear differential equation (\ref{1})}
 if the minimum $\min_{i,j} \{a_i^{(j)}+Val_{S_i} (j),\, a\}$ is attained at least twice or is infinite (as it is
accustomed in tropical mathematics \cite{I}, \cite{M}). The latter
is a necessary condition of solvability in power series in $t$ of a
linear differential equation $\sum _{i,j} A_{i,j}\cdot X_i^{(j)}=A$
in several indeterminates $X_1,\dots,X_n$. Namely, the orders of
power series coefficients equal $ord_t (A_{i,j})=a_{i,j},\, ord_t
(A)=a$ and the support of $X_i$ is $S_i$. More precisely, (\ref{1})
expresses that at least two lowest terms of the expansion in power
series of the differential equation have the same exponents, which
is similar to that the tropical equations concern the lowest terms
of the
 expansions in Puiseux series of algebraic equations.

We study solvability of a system of tropical linear differential equations
\begin{eqnarray}\label{2}
\min_{i,j} \{a_{i,l}^{(j)} + x_i^{(j)},\, a_l\},\, 1\le l\le k
\end{eqnarray}
where $1\le i\le n,\, 0\le j\le r$ and for all finite coefficients $a_{i,l}^{(j)},\, a_l \in \ZZ$
we have $0\le a_{i,l}^{(j)},\, a_l \le M$. Thus, the bit-size of (\ref{2}) is bounded by
$knr\log_2 (M+2)$.

We say that a solution $T_1,\dots, T_n$ of (\ref{2}) is {\it minimal} if the inequality $Val_{T_1,\dots,
T_n} \le Val_{S_1,\dots,S_n}$ holds pointwise for any solution $S_1,\dots,S_n$ of  (\ref{2}).

Note that  (\ref{2}) extends tropical linear systems when for all the occurring derivatives $x_i^{(j)}$
we have $j=0$. Thus, the complexity bound of testing solvability of  (\ref{2}) in the next theorem
generalizes the similar complexity bound of solvability of tropical linear systems from \cite{A}, \cite{B},
\cite{G}.

\begin{theorem}\label{main}
If a system  (\ref{2}) of tropical linear differential equations is solvable then it has the (unique)
minimal solution. There is an algorithm which tests solvability of  (\ref{2}) and in case of solvability
yields its minimal solution within the complexity polynomial in $knrM$.
\end{theorem}

Note that $S\subset T \subset \ZZ_+$ iff the inequality $Val_S \ge
Val_T$ holds pointwise. For $S_1,\dots,S_n,T_1,\dots,T_n \subset
\ZZ_+$ we have for the pointwise minimum $$Val_{(S_1,\dots,S_n) \vee
(T_1,\dots,T_n)} := Val_{S_1\cup T_1,\dots S_n \cup T_n} = \min
\{Val_{S_1,\dots,S_n},\, Val_{T_1,\dots,T_n}\}.$$

Assume now that (\ref{2}) has a solution $S_1,\dots,S_n$. If $s\in
S_i$ such that $s\ge r$ then one can replace $S_i$ by adding to it
all the integers greater than $s$ (while keeping $S_1,\dots,S_n$
still to be a solution of (\ref{2})). Therefore, one can suppose
w.l.o.g. that for every $1\le i\le n$ either $S_i$ is finite and
moreover $S_i \subset \{0,\dots,r-1\}$ or the complement $\ZZ_+
\setminus S_i$ is finite. Thus, if we define $V:= Val_{\bigvee
(S_1,\dots,S_n)}$ where $\bigvee$ ranges over all the solutions
$S_1,\dots,S_n$ of (\ref{2}) then $\bigvee$ can be taken over a
finite number of solutions, hence it can be reduced to a single
solution $T_1,\dots,T_n$, thereby $V=Val_{T_1,\dots,T_n}$ and
$T_1,\dots,T_n$ is the minimal solution of (\ref{2}) which proves
the first statement of Theorem~\ref{main}.

Also we design a polynomial complexity algorithm for solving systems
of the type (\ref{1}) in case of one variable ($n=1$).

\begin{theorem}\label{one}
There is an algorithm which tests solvability of a system (\ref{1})
of tropical linear differential equations in one variable $x$ and
yields its minimal solution in case of solvability within the
polynomial complexity. More precisely,  the complexity is bounded by
$O(kr\log(rM))$.
\end{theorem}

\section{Bound on the minimal solution of a tropical linear
differential equation}\label{estimate}

Our next goal is to bound the (finite) complements $\ZZ \setminus
S_i$. For each $1\le i\le n$ with a finite $\ZZ \setminus S_i$
denote by $m_i \in S_i$ the minimal element of $S_i$ such that $m_i
\ge r$. If for some $1\le l\le k$ the inequality $\min_{i,j}
\{a_{i,l}^{(j)} + Val_{S_i} (j),\, a_l\} > M+r$ holds then for every
$1\le i_0\le n,\, 0\le j_0 \le r$ for which this minimum is
attained: $a_{i_0,l}^{(j_0)} + Val_{S_{i_0}} (j_0) = \min_{i,j}
\{a_{i,l}^{(j)} + Val_{S_i} (j),\, a_l\}$ we have $Val_{S_{i_0}}
(j_0) = m_{i_0}-j_0$.

Consider a graph $G$ which for each finite $S_i, 1\le i\le n$
contains a vertex $w_i$ and for each $S_i$ with a finite complement
$\ZZ_+ \setminus S_i$ contains two vertices $w_i,\, w_i^{\infty}$. A
derivative $x_i^{(j)}$ corresponds to a vertex $w_i^{\infty}$ iff
$Val_{S_i} (j) = m_i-j$ (provided that $\ZZ_+ \setminus S_i$ is
finite), else $x_i^{(j)}$ corresponds to $w_i$. Also $G$ contains a
vertex $w_0$ to which corresponds every free term $a_l, 1\le l\le
k$.

If there are $1\le l\le k,\, 1\le i_0,i_1\le n,\, 0\le j_0,j_1 \le
r$ such that
\begin{eqnarray}\label{3}
a_{i_0,l}^{(j_0)}+ Val_{S_{i_0}} (j_0) = a_{i_1,l}^{(j_1)}+
Val_{S_{i_1}} (j_1) = \min_{i,j} \{a_{i,l}^{(j)}+ Val_{S_{i}} (j),\,
a_l\} < \infty
\end{eqnarray}
\noindent then we connect in $G$ by an edge vertices which
correspond to the derivatives $x_{i_0}^{(j_0)}$ and
$x_{i_1}^{(j_1)}$. Instead of $a_{i_0,l}^{(j_0)}+ Val_{S_{i_0}}
(j_0)$ could be $a_l$, then we consider the vertex $w_0$.

If a connected component of $G$ contains only vertices of the form
$w_i^{\infty}$ with $m_i>r$ then for each $w_i^{\infty}$ from this
component we replace $m_i$ by $m_i-1$, or in other terms, augment
$S_i$ by $m_i-1$, preserving so modified $S_1,\dots,S_n$ (for which
we keep the same notation) to be still a solution of (\ref{2}).
After that $G$ can be modified (we use the same notation for the
modified $G$), and we continue this process. Eventually, we arrive
to a solution $S_1,\dots,S_n$ whose graph $G$ has no connected
component satisfying the described property.

Therefore, each connected component of $G$ contains a vertex of the
form $w_{i_0}$ (or perhaps, $w_{i_0}^{\infty}$ with $m_{i_0}=r$)
fulfilling (\ref{3}) for suitable $j_0,\,l$. Then $Val_{S_{i_0}}
(j_0) \le r$ (there is a possibility that instead of
$a_{i_0,l}^{(j_0)}+ Val_{S_{i_0}} (j_0)$ we consider the free term
$a_l$), hence $Val_{S_1} (j_1)\le M+r$ follows from (\ref{3}). Thus,
for every vertex $w_i^{\infty}$ from this connected component there
is a path in $G$ of a length at most $n-1$ connecting it with a
vertex of the form $w_{i_2}$ (or perhaps, $w_{i_2}^{\infty}$ with
$m_{i_2}=r$). Therefore, there is $0\le j\le r$ such that
$m_i-j=Val_{S_i} (j)\le (n-1)(M+r)$ (one can show the latter
inequality following along the path and applying the above
argument). Thus, we conclude with the following lemma.

\begin{lemma}\label{bound}
For each $1\le i\le n$ for which $\ZZ_+ \setminus S_i$ is finite the
bound  $m_i \le N:= (n-1)(M+r)+r$ holds.
\end{lemma}

This lemma extends Lemmas 1.2, 2.2 \cite{G} established for tropical
linear systems.

\section{Algorithm testing solvability and producing the minimal
solution of a system of tropical linear differential
equations}\label{algorithm}

Now we proceed to design an algorithm which tests whether a system (\ref{2})
is solvable and if yes then yields its minimal solution $T_1,\dots,T_n$. The
algorithm starts with the setting $T_1=\cdots=T_n=\{0,\dots,N\}$ (see Lemma~\ref{bound}),
perhaps, being not a solution of (\ref{2}), and then modifies $T_1,\dots,T_n$
recursively while a current $T_1,\dots,T_n$ is not a solution. If eventually a current
$T_1,\dots,T_n$ becomes a solution then it is the minimal solution. We show by recursion
that for any solution $S_1,\dots,S_n$ of (\ref{2}) the pointwise inequality
$Val_{S_1,\dots,S_n} \ge Val_{T_1,\dots,T_n}$ holds for a current $T_1,\dots,T_n$.

If a current $T_1,\dots,T_n$ is not a solution of (\ref{2}) then two
cases can emerge. In the first case there exist $1\le i_0 \le n,\,
0\le j_0\le r,\, 1\le l\le k$ such that a finite minimum $\min_{i,j}
\{a_{i,l}^{(j)}+Val_{T_i}(j)\} < a_l$ is attained at a single pair
$i_0,\, j_0$. Let $Val_{T_{i_0}} (j_0)=s-j_0$ where $s\in T_{i_0}$
is the minimal element of $T_{i_0}$ such that $s\ge j_0$. The
algorithm modifies $T_{i_0}$ discarding $s$ from it. The inequality
$Val_{S_1,\dots,S_n} \ge Val_{T_1,\dots,T_n}$ still holds for any
solution
 $S_1,\dots,S_n$ of (\ref{2}) since $S_i\subset T_i,\, 1\le i\le n$. Note that if $s=m_{i_0}<N$
 (see Lemma~\ref{bound}) then $m_{i_0}$ increases by one. If $s=N(=m_{i_0})$ then $S_{i_0}$ is
finite due to Lemma~\ref{bound} and the algorithm discards from the current $T_{i_0}$ all integers
$p>N$.
In the second case $a_l$ is the unique minimum in $\min_{i,j} \{a_{i,l}^{(j)}+Val_{T_i}(j),\, a_l\}$,
then system (\ref{2}) has no solution.

The algorithm terminates when either $T_1,\dots,T_n$ is a solution of (\ref{2}), in this
case $T_1,\dots,T_n$ is the minimal solution, or the algorithm detects that (\ref{2}) has no solution.

When (\ref{2}) is homogeneous, i.~e. $a_l=\infty,\, 1\le l\le k$, system (\ref{2}) has a solution
with all infinite functions $Val_{S_i},\, 1\le i\le n$. It can happen that the algorithm terminates
with all $T_1,\dots,T_n$ being void, that means that the infinite solution of (\ref{2}) is its
unique one.

To bound the complexity of the algorithm observe that it runs at
most $n(N+1)$ steps because at each step at least one of the current
sets $T_1,\dots,T_n \subset \{0,\dots,N\}$ decreases. The cost of
each step is polynomial in $knr\log M$ (the algorithm for every
$1\le i\le n$ stores $m_i$, provided that $\ZZ_+ \setminus T_i$ is
finite, and also stores $T_i\cap \{0,\dots,r-1\}$). This completes
the proof of Theorem~\ref{main}.

\section{Polynomial complexity solving systems of tropical linear
differential equations in one variable}\label{polynomial}

We design a polynomial complexity algorithm for solving a system of
tropical linear differential equations (\ref{2}) in one variable
$x$. The algorithm basically follows the algorithm from
Theorem~\ref{main} with a few modifications. In fact, the algorithm
designed in this Section is a version of the algorithm from
Theorem~\ref{main}, the modification consists in that its steps are
ordered in a special way (observe that at each step of the algorithm
from Theorem~\ref{main} there could be several choices of an element
to be discarded from the current set $T$). We use the notations from
Section~\ref{algorithm}.

First, if there exists $s<r$ from  $T$ such that a finite minimum
$\min_j \{a_l^{(j)} + Val_T (j)\}$ is attained at a unique $j_0$ for
some $1\le l\le k$ and it holds $Val_T(j_0)=s-j_0$ then the
algorithm discards $s$ from $T$. This is also a step of the
algorithm from Theorem~\ref{main}, and we refer to it as a {\it step
of the finite type}. In other words, the algorithm designed in this
Section has a preference in discarding elements $s$ which are less
than $r$. Denote by $s_0\ge r$ the minimal element of $T\cap
[r,\infty)$.

Second, let otherwise $s_0$  be the only candidate to be discarded
from $T$ for all the equations from (\ref{2}) which are not
satisfied by $T$. Then the algorithm from Theorem~\ref{main} would
just discard $s_0$, while the algorithm under description discards
from $T$ possibly more elements at one step.

For each $1\le l\le k$ consider a unique $0\le j_0\le r$ (provided
that it does exist, i.~e. the $l$-th equation is not satisfied by
$T$) such that $Val_T (j_0)=s_0-j_0$ and $a_l^{(j_0)}+s_0-j_0=
\min_j \{a_l^{(j)} + Val_T (j)\}$. Take the maximal $p_l$ (or the
infinity when it is not defined) such that \begin{eqnarray}\label{8}
a_l^{(j_0)}+s_0-j_0+p_l \le a_l^{(j)}+Val_T(j) \end{eqnarray} for
any $0\le j<r$ for which $Val_T(j)=s_1-j$ for suitable $T\ni s_1<r$.
Observe that $p_l\ge 1$.

Denote by $p$ the maximum of all such $p_l$. Let $p=p_{l_0}$ for an
appropriate $1\le l_0\le k$. If $p=\infty$ then the algorithm
discards from $T$ all the elements $s\ge r$. Else, if $p< \infty$
then the algorithm discards all the elements $s$ from $T$ such that
$s_0\le s < s_0+p$. In other words, the algorithm replaces the
minimal element $s_0$ of $T\cap [r,\infty)$ by $s_0+p$, we call this
step of the algorithm a {\it jump}. Clearly, the jump replaces $p$
steps of the algorithm from Theorem~\ref{main} each consisting in
discarding just one element from $T$ (so, discarding consecutively
$s_0,s_0+1,\dots,s_0+p-1$) due to the unique monomial
$a_{l_0}^{(j_0)} + x^{(j_0)}$ at which the minimum in (\ref{2}) is
attained for the $l_0$-th equation.

Observe that after a jump either the (new current) $T$ provides a
solution of (\ref{2}) or the algorithm can execute a step of the
finite type because of the choice of $p$, see (\ref{8}), so discards
from $T$ some element $s<r$.

As in Section~\ref{algorithm} the algorithm terminates when either a
current $T$ provides a solution of (\ref{2}) (being the minimal
solution as it was proved in Section~\ref{algorithm}) or the
algorithm exhausts $T$ (which means that $T\cap [0,N]=\emptyset$,
see Sections~\ref{estimate}, \ref{algorithm}). In the latter case if
system (\ref{2}) is homogeneous then it has the (unique) infinite
solution, otherwise a non-homogeneous system has no solutions (again
similar to Section~\ref{algorithm}).

To estimate the complexity of the algorithm note that  after a jump
the algorithm executes a step of the finite type, i.~e. discards
from $T$ an element $s<r$. Therefore, the number of steps of the
algorithm does not exceed $2r$ taking into the account that the
number of steps of the finite type is less or equal than $r$. To
bound the jump $p$ observe that
$a_{l_0}^{(j_0)}+s_0-j_0+p=a_{l_0}^{(j_1)} +s_1-j_1$ for appropriate
$j_1,s_1<r$  (cf. (\ref{8})). Since $j_0\le r$ we deduce that
$s_0+p< 2r+M$, hence $p<r+M$. Thus, one can estimate the complexity
by $O(kr\log(rm))$, and we complete the proof of Theorem~\ref{one}.

\section{$NP$-hardness of solvability of tropical non-linear
differential equations in one variable}

Now generalizing tropical linear differential equations (see the
Introduction) we consider systems of tropical {\it non-linear}
differential equations of the form
\begin{eqnarray}\label{4} \min_{\{P\}} \{a_P+\sum _{(i,j)\in P}
x_i^{(j)}\} \end{eqnarray} where the coefficients $a_P\in \ZZ_+$ and
the minimum ranges over a certain (finite) family of finite
multisets $P$ of pairs $(i,j)$. We view $|P|$ as the degree of the
monomial $a_P+\sum _{(i,j)\in P} x_i^{(j)}$.

Similar to the case of tropical linear differential equations (see
the Introduction), we observe that the solvability of (\ref{4}) is
necessary for the solvability in power series in $t$ of a non-linear
differential equation $\sum_{\{P\}} A_P \cdot \prod _{(i,j)\in P}
X_i^{(j)} =0$ where $ord_t (A_P)=a_P$.

We prove that the problem of solvability (with a set $S\subset
\ZZ_+$ similar to tropical linear differential equations, see the
Introduction) of a system of equations of the form (\ref{4}) is
$NP$-hard already in the case of a single variable $x$. Mention that
in \cite{T} $NP$-completeness of the solvability of tropical
non-linear systems (in several variables) is established.

We prove $NP$-hardness by means of reducing a 3-$SAT$ boolean
formula $\Phi$ in $n$ variables $y_0,\dots,y_{n-1}$ (see e.~g.
\cite{J}) to a system $E_{\Phi}$ of equations of the form (\ref{4})
in a single variable $x$, preserving the property of solvability.

The system $E_{\Phi}$ contains (linear) equations
\begin{eqnarray}\label{5} \min \{x^{(2j+1)},\, 0\},\, 0\le j\le
2n-1 \end{eqnarray} These equations mean that the valuation of each
even derivative $x^{(2j)},\, 0\le j\le 2n-1$ equals either $0$ or
$1$. Also $E_{\Phi}$ contains (quadratic) equations
\begin{eqnarray}\label{6} \min \{x^{(2j)}+x^{(2j+2n)},\, 1\},\, 0\le
j\le n-1 \end{eqnarray} They mean that either $Val(x^{(2j)})=0,\,
Val(x^{(2j+2n)})=1$ (which corresponds to the value "true" of the
variable $y_j,\, 0\le j\le n-1$ of $\Phi$) or $Val(x^{(2j)})=1,\,
Val(x^{(2j+2n)})=0$ (which corresponds to the value "false" of
$y_j$, respectively). Finally, for each 3-clause of $\Phi$, say of
the form $\neg y_{j_1} \vee y_{j_2} \vee \neg y_{j_3}$ we add to
$E_{\Phi}$ a (linear) equation \begin{eqnarray}\label{7} \min
\{x^{(2j_1+2n)},\, x^{(2j_2)},\, x^{(2j_3+2n)},\, 0\} \end{eqnarray}
Clearly, $\Phi$ is equivalent to the solvability of the system
obtained by uniting (\ref{5}), (\ref{6}) and (\ref{7}) for all
3-clauses of $\Phi$. Thus, we have proved

\begin{proposition}
The problem of solvability of systems of tropical non-linear
differential equations in a single variable is $NP$-hard.
\end{proposition}

\section{Solvability of systems of tropical non-linear differential
equations is in $NP$}

Next we prove that the problem of solvability of systems of $k$
tropical non-linear differential equations of the form (\ref{4}) of
degrees $|P|\le d$ fulfilling the bounds: $0\le a_P\le M,\, 0\le
j\le r,\, 1\le i\le n$ (in an arbitrary number $n$ of variables)
belongs to $NP$. First, similar to Lemma~\ref{bound} and using the
notations from Section~\ref{estimate}, we show that if a system has
a solution (with some $S_1,\dots,S_n \subset \ZZ_+$) then it
possesses a sufficiently small solution.

Substitute the solution into each equation of the form (\ref{4}),
then the valuations of some derivatives $x_i^{(j)}$ can equal
$m_i-j$, the valuations of all the other derivatives consider as
being fixed. We treat the system (after this substitution) as an
input of the linear programming problem (expressing that the minimum
in (\ref{4}) is attained at least at two terms) with respect to the
indeterminates $m_i$ (for all $i$ for which they are defined), and
the fixed valuations consider as the coefficients of the input.
Therefore, this input possesses a solution with $m_i$ bounded (due
to Hadamard's inequality on determinants) by $N_1:= n!\cdot
(M+rd)\cdot d^n$. Note that this bound is worse than the bound on
$N$ established in Lemma~\ref{bound} for systems of tropical {\it
linear} differential equations.

Since in order to give a solution $S_1,\dots,S_n$ it suffices just
to specify $S_i \cap [0,r],\, 1\le i\le n$ and $m_i\le N_1$ (for $i$
for which it does exist), we get the following

\begin{proposition}\label{np}
The problem of solvability of systems of tropical non-linear
differential equations belongs to $NP$.
\end{proposition}

\section{Further research}

Similar to tropical linear systems (cf. \cite{A}, \cite{B},
\cite{G}) it is an open problem, whether one can solve system
(\ref{2}) of tropical linear differential equations within the
complexity polynomial in $knr\log M$ (in other words, within the
proper polynomial complexity)? In Section~\ref{polynomial} a
polynomial complexity algorithm is designed for testing solvability
of systems of tropical linear differential equations in one variable
($n=1$). Is there a polynomial complexity algorithm for similar
systems in, say a constant number $n\ge 2$ of variables?

It is known (see \cite{A}, \cite{B}, \cite{G}) that the problem of
solvability of systems of tropical linear equations is in the
complexity class $NP \cap coNP$. Does the problem of solvability of
systems of the type (\ref{2}) of tropical linear differential
equations belong to $coNP$? Proposition~\ref{np} implies that even a
more general problem of solvability of systems of the type (\ref{4})
of tropical non-linear differential equations lies in $NP$.

It is proved in \cite{SS} the following coincidence for the closure
in the euclidean topology: $\overline{Trop (V(I))}=V(Trop(I))\subset
\RR^n$ where $I\subset K[X_1,\dots,X_n]$ is a polynomial ideal over
the field $K$ of Puiseux series and $V(I)\subset K^n$ is the variety
of $I$. Does there hold an analogue of this coincidence for
differential ideals? In other words, is it true that for any
differential ideal $G$ in $n$ independent variables and a family
$S_1,\dots,S_n\subset \ZZ_+$ being a solution of the tropical
differential equation $Trop(g)$ for any $g\in G$, there exists a
power series solution of $G$ whose tropicalization equals
$S_1,\dots,S_n$?

We say that $S_1,\dots,S_n$ is a {\it Laurent solution} of (\ref{2})
if for every $1\le i\le n$ either $S_i \subset \ZZ_+$ is as we
considered above or $S_i=\{b\}$ is a singleton for some negative
integer $0>b\in \ZZ$. In the latter case $Val_b(j)=b-j$. This
corresponds to the order of the $j$-th derivative of a Laurent
polynomial of the form $ct^b(1+O(t))$ for a (complex) coefficient
$c$. If all sets among $S_1,\dots,S_n$ are negative singletons then
the solvability of (\ref{2}) reduces to the solvability of a
tropical linear system. The question is, what is the complexity of
testing whether (\ref{2}) has a Laurent solution? Actually, one can
extend this setting from Laurent solutions to solutions of the form
$S_i=\{b\}$ where $b\in \RR \setminus \ZZ_+$. This corresponds to a
necessary condition of solvability of a system of linear
differential equations in Puiseux series (when $b\in \QQ$) or in
Hahn series (when $b\in \RR$, see e.~g. \cite{S}).


For a tropical linear differential monomial $a+x^{(j)},\, a,j \in
\ZZ_+$ define its derivative as $\min\{a-1+ x^{(j)},\,
a+x^{(j+1)}\}$ when $a\ge 1$ or as $x^{(j+1)}$ when $a=0$ (which
mimics the usual derivation law). We spread this definition of the
derivative to all tropical linear differential equations of the form
(\ref{2}) by the tropical linearity. The tropical ideal generated by
the derivatives of all the orders of tropical linear differential
equations is called the {\it tropical linear differential ideal}
generated by these equations. Is it possible to test solvability of
a tropical linear differential ideal? Lest there would be a
misunderstanding, we note that a solution of a tropical linear
differential equation is not necessary a solution of the tropical
ideal generated by this equation. \vspace{2mm}

{\bf Acknowledgements}. The authors is grateful to the Max-Planck
Institut f\"ur Mathematik, Bonn for its hospitality during writing
this paper.

\end{document}